# Further correcting pressure effects on SBE911 CTD-conductivity data from hadal depths

by Hans van Haren[1*], Hiroshi Uchida[2], Daigo Yanagimoto[3]

[1]Royal Netherlands Institute for Sea Research (NIOZ) and Utrecht University, P.O. Box 59, 1790 AB Den Burg, the Netherlands.
*e-mail: hans.van.haren@nioz.nl

[2] Research Institute for Global Change, Japan Agency for Marine-Earth Science and Technology (JAMSTEC), Yokosuka, Kanagawa, Japan
[3]Atmosphere and Ocean Research Institute, The University of Tokyo
Kashiwanoha 5-1-5, Kashiwa, Chiba 277-8564, Japan

**Abstract**

Hadal, >6000 m deep shipborne Sea-Bird Electronics SBE 911plus Conductivity Temperature Depth (CTD) data are obtained using two different systems in the vicinity of the Earth's deepest point, in the Challenger Deep, Mariana Trench 14 years apart. Below 7000 m in very weakly density stratified waters, the salinity data from both sets show an artificial increase with depth of about $10^{-6}$ g kg$^{-1}$ m$^{-1}$ that is not covered by the SBE linear pressure correction to the conductivity data. With the aid of independent water sample data, salinity is corrected and an additional algorithm for the pressure correction on conductivity data is formulated. The corrected salinity data still weakly increase with depth, together with a decrease in temperature, which may point at an influx of dense modified Antarctic bottom water. The corrected density variations with depth are used in calculations of deep turbulence values.

**1 Introduction**

Due to the large hydrostatic ambient pressure, few observations on deep-sea life conditions have been made in hadal zones deeper than 6000 m so far. Interesting marine biological questions are to be addressed on life under such harsh conditions, questions varying from molecular cellular level to sparsely distributed ecological communities both in the water column (e.g., Jamieson, 2015; Gallo et al. 2015; Nunoura et al. 2015) and in sediments of deep-ocean floors (e.g., Glud et al. 2013). Dynamical variations in the physical environment are also not well known, except that hadal waters in ocean trenches cannot be stagnant pools of cold water as species like crustaceans require sufficient replenishment of oxygen and nutrients (Johnson 1998). Further quantitative knowledge is thus required on transport processes, which are likely driven by density differences on the large scales and turbulent overturning motions on the small scales. Although progress is slow because of the required specialized instrumentation and long cables for lowering shipborne equipment, several attempts have improved the electronic sampling in the last decades (e.g., Kawagucci et al. 2018).

Water mass characteristics like temperature and salinity determining density variations were the first static physical properties to be studied. Before the use of electronic devices, discrete inverse thermometer readings were made from R/V Vityaz in the late 1950's and a small free-falling water-sampling device equipped with reversing thermometers and water samplers was dropped to the floor of the Mariana Trenches' Challenger Deep in 1976 (Mantyla and Reid 1978). The water sample data were analyzed in the laboratory for the salinity content that appeared constant to within 0.001 ppt. These data were used by Taira et al. (2005) as a reference for their deep shipborne Sea-Bird Electronics SBE-911 Conductivity Temperature Depth CTD-cast attached to a custom-made titanium wire, which was lowered down to 10877 m using a custom-made swell-compensator winch. Taira et al.'s (2005) findings led to a linear pressure-correction of the conductivity data due to length and diameter changes in the borosilicate glass cell as a function of pressure that became apparent at great depths (Sea-Bird Electronics 2013). However, even after accomplishing such correction, an additional artificial pressure effect remained in the data from recent casts into the Challenger Deep reaching a depth of 10851 m in 10907 m water depth (van Haren et al. 2017). The resulting artificial increase in salinity and thus density not only hampers water mass analyses in hadal zones, but potentially also computations on turbulence.

As turbulence microstructure profilers and (lowered) acoustic Doppler current profilers for measuring shear variance (e.g., Kunze et al. 2006) do not go deeper than 6000 m to date, we rely on shipborne CTD-instrumentation to calculate turbulent overturning displacements and values on turbulence dissipation rate and vertical eddy diffusivity following the method proposed by Thorpe (1977). However, the standard high-resolution SBE-instrumentation has to become modified in several ways to be able to calculate turbulence values within acceptable error bounds of about a factor of three. An important modification is to minimize the effects of surface-wave-induced ship motions (Taira et al. 2005) and rotation and related motions of the instrument-package (Uchida et al. 2015) on the electronic CTD-data. A direct way of minimizing is the use of an effective swell-compensator (Taira et al. 2005) together with an underwater slip ring swivel and a stabilizing fin (Uhida et al. 2018). Alternatively, during the post-processing the effects of swell may be removed via low-pass filtering the recorded raw data (van Haren 2015).

In this note, we report on a mathematical method to correct for an additional artificial pressure effect in deep high-resolution SBE-911 CTD casts into the Challenger Deep. One data set was obtained together with water sample data. This served as a lead for the proposed nonlinear mathematical correction. The correction is subsequently applied to the Challenger Deep data of van Haren et al. (2017) that serve as an example for other hadal-deep SBE-911 CTD-data. Our primary interest is in finding the approximately correct vertical slopes of CTD-profiles, culminating in the most realistic vertical slope of density anomaly values. Such slopes express the amount of vertical density variation 'stratification' and the work of turbulence kinetic energy against it. Hereby we do not worry about the exact absolute values as obtained from calibrations and used in water mass definitions and studies on variations therein.

**2 Observations and data**

Shipborne CTD observations have been made from the Japanese R/V Hakuho Maru in August 2002 and the German R/V Sonne in November 2016 above two deep spots of the Challenger Deep, the southernmost part of the Mariana Trench including the world's deepest point (Fig. 1). In both cases, a Sea-Bird SBE911plus CTD was used and the instrumentation package was mounted in a horizontal position at the bottom of a water sampler frame.

The Hakuho Maru lowered the CTD using a custom-made titanium cable with deepest measurement 10366 m at 11° 21.0′N, 142° 24.5′ E in 10346±25 m water depth. To minimize ship-motion effects on the CTD-measurements, an active swell compensator, manufactured by Mitsubishi Heavy Industries, was controlled by an integrated signal of the accelerometer mounted on a vertical gyroscope under the gallows (Taira et al. 2005). In spectra of the principle parameter data swell and wind-wave peak were indeed absent. It is noted that in the surface wave range, for frequencies exceeding 0.02 cps, cycles per second, the spectra of deep CTD-data were flat white noise rather than continuing the slow decrease with frequency as at lower frequencies. During the CTD-upcast, water samples were taken for laboratory salinity determination by salinometer (Portasal 8410A, Guildline Instruments Ltd., Ontario, Canada). In hindsight, in the 2002-CTD the temperature sensor (SBE3 S/N893) had an abnormal pressure hysteresis of up to about ±0.001ºC. To correct it, the pressure dependency was estimated from comparisons with SBE35 and another SBE3 (S/N5982) during a following cruise (Kawagucci et al. 2018).

The Sonne lowered the CTD with freshly calibrated T-C sensors using its 18 mm steel cable at 11° 19.752′N, 142° 11.277′ E in 10907±12 m water depth with deepest measurement 10851 m (van Haren et al. 2017). The ship motion effects were removed during post-processing by applying a sharp double elliptic, phase-preserving low-pass filter (Parks and Burrus 1987) with 0.05 cps, cycles per second, cut-off frequency. For consistency between the data sets, the Hakuho Maru data are also post-processed with this low-pass filter to remove some of the white noise. No water samples were taken during the Sonne's CTD-upcast as the Rosette-electronic bottle firing system was not rated at pressures exceeding 6000 dbar.

Following the Taira et al. (2005) publication, Sea-Bird Electronics introduced an extra pressure correction on conductivity data that accommodated for the small linear changes in ceramics cell-length under static pressure (Sea-Bird Electronics 2013),

$$C(0) = C(p)/(1 + a \cdot p) \quad [S\ m^{-1}], \tag{1}$$

in which C denotes conductivity (at pressure p) and $a = -9.57 \times 10^{-8}$ [dbar$^{-1}$] the nominal constant correction factor. With this correction, the full calibration equation from raw frequency (f) data reads (e.g., Sea-Bird Electronics 2013),

$$C(p, T) = (g + h \cdot f^2 + i \cdot f^3 + j \cdot f^4)/10(1 + b \cdot T + a \cdot p) \quad [S\ m^{-1}], \tag{2}$$

where g,h,i and j are calibration constants and b is a constant for linear temperature T correction. Equation (2) becomes nonlinear in p when temperature is a nonlinear function of pressure.

We processed the CTD-data using the standard procedures incorporated in the SBE-software, including corrections for cell thermal mass using the parameter setting of Mensah et al. (2009) and sensor time-alignment. Conservative (~potential) Temperature ($\Theta$), absolute salinity SA and density anomalies $\sigma_{11}$ referenced to 11000 dbar are calculated using the Gibbs-SeaWater GSW-software that define thermodynamic properties of seawater based on a Gibbs function formulation (IOC, SCOR, IAPSO 2010).

Estimates of turbulence dissipation rate $\varepsilon_T = c_1^2 d^2 N^3$ and vertical eddy diffusivity $K_{zT} = m_1 c_1^2 d^2 N$ are made from the CTD-downcast data using the method of reordering potentially unstable vertical density profiles in statically stable ones, as proposed by Thorpe (1977). Here, d denotes the displacements between unordered (measured) and reordered profiles. N denotes the buoyancy frequency computed from the reordered profiles. Rms-values of displacements are not determined over individual overturns, as in Dillon (1982), but over 200 m vertical intervals that just exceed the largest overturn intervals. This avoids the complex distinction of smaller overturns in larger ones and allows the use of a single averaging length scale. We use standard constant values of $c_1 = 0.8$ for the Ozmidov/overturn scale factor (Dillon 1982) and $m_1$ = 0.2 for the mixing efficiency (Osborn 1980; Oakey 1982). This is the most commonly used parameterization for oceanographic data, see further discussions in van Haren et al. (2017) and Gregg et al. (2018). For eddy diffusivity also other parametrizations are used, for comparison. As a criterion for determining overturns from the surface wave corrected density data, we only used those data of which the absolute value of difference with the local reordered value exceeds a threshold of $7\times10^{-5}$ kg m$^{-3}$, which corresponds to applying a threshold of $1.4\times10^{-3}$ kg m$^{-3}$ to raw data variations (e.g., Galbraith and Kelley 1996; Stansfield et al. 2001; Gargett and Garner 2008).

**3 Results and further correction**

**3.1 General CTD-profiles obtained from Challenger Deep 14 years apart**

The SBE-software processed and ship-motion corrected deeper half of the 2002 and 2016 CTD-profiles in the Challenger Deep are generally consistent (Fig. 2). Modifications on the large scale between the profiles, like a temperature difference of about 0.005°C and a salinity difference of about 0.0015 g kg$^{-1}$, are considered due to water mass variations between the times of data taking, 14 years apart, and, on a shorter time scale, due to internal wave action, mesoscale eddies and sub-mesoscale processes like occurring near the ocean-surface (e.g., Nakano et al. 2015; Qiu et al. 2020). Pressure dependency (Uchida et al. 2015) of the temperature sensors and the batch-to-batch differences of the standard seawater (Uchida et al. 2020) used in the manufacturer's calibration for the conductivity sensors may also contribute to the differences.

Around z = -6000 m at the top of the trench or the level of the surrounding ocean floor, the vertical variation in temperature, salinity and density becomes very low. In Conservative Temperature (Fig. 2a), a near-homogeneous layer is observed between -8300 < z < -7000 m in the 2002-profile.The 2002-salinity profile demonstrates a large-scale weak instability between -7500 < z < -5500 m that is not observed in the 2016-Sonne CTD-downcast data (Fig. 2b). It is a weaker version of the S-shape instability in the data presented by Taira et al. (2005), who used the same serial numbered CTD-sensors as during the 2002-Hakuho-Maru cruise, and attributable to the linear pressure effects to the ceramics of the conductivity cell that are corrected following (1) since SeaBird Electronics (2013).

Of concern here is the exponential increase with depth in salinity for $z < -7500$ m. Despite this exponential increase being found consistently in all profiles we consider it artificial.

The above observations in salinity are reflected in the density anomaly data (Fig. 2c). It is noted that for $z < -8500$ m the 2002- and 2016-profiles almost overlap, which indicates an almost complete matching of different salinity and temperature contributions to density variations.

In both data sets, small oscillations of about 0.0001ºC are visible especially in the weakly stratified layer between about $-9000 < z < -7000$ m. These small oscillations are considered realistic and typical for deep trench instabilities or overturning. The oscillations are part of 100 to 200 m tall overturns, with a general shape of large-scale weak stability above a small-scale relatively strong instability (Fig. 3).

**3.2 Finding a correction for artificial conductivity data**

The 2002-salinity-data are corrected using water sample data analysed in the laboratory. Although the individual water sample data showed considerable spike behaviour of 0.001 g kg$^{-1}$ and larger, the general trend shows a weakly, not-exponentially increasing profile with depth (Fig. 4). The correction applied is thus one that yields a more or less steady weakly stable vertical gradient in salinity for $z < -6700$ m, leaving the small-100-m-scale instabilities unaffected.

In order to apply the above salinity correction to other deep hadal CTD-profiles we recomputed a corrected conductivity profile $C_c(p)$ from the corrected salinity, corrected temperature and pressure 2002-CTD-data and compared it with the original conductivity profile $C_o(p)$ from (2) (Fig. 5a). The small jump in conductivity difference $\Delta C = C_c - C_o$ of about 0.00001 S m$^{-1}$ around $z = -7540$ m was caused by the temperature correction mentioned in Section 2.

We fitted the conductivity difference data with a 9$^{th}$ order polynomial on pressure. The coefficients are given in the Appendix. This high order was required to more or less correctly grasp the entire conductivity difference profile from surface to trench floor, due to some wild variations in conductivity related with water property variations near the surface. Alternatively, we fitted a lower 3$^{rd}$ order polynomial that provided the same standard deviation between fit and $\Delta C$, but only covering z < -5000 m deep waters, which we find less preferential. We also attempted to fit a polynomial to corrected salinity data difference with original data, as a function of z, but the results were less good (in salinity and density profiles) than fitting the $\Delta C(p)$.

The difference in salinity between the 9$^{th}$-order polynomial fit and the water sample corrections is smaller than $\pm 2 \times 10^{-4}$ g kg$^{-1}$ (Fig. 5b). It is one order of magnitude smaller than the differences between electronic and laboratory water sample data and the salinity difference between original 2016-Sonne data and the 9$^{th}$-order polynomial correction to their conductivity data using (A1). In vertical salinity profiles for z < -6000 m, the corrections show a slight weakening of the general vertical salinity gradient for the 2002-CTD-data and a larger vertical salinity gradient in the 2016-CTD-data that is steady on the 1000-m scale (Fig. 5c).

**3.3 Turbulence calculations**

In density anomaly, the two deep profiles have different slopes with depth that become near-equal for z < -9500 m (Fig. 6a). In this deep range, the effects of the corrected conductivity data on computed turbulence values is evident but smaller than expected.

Between -6000 < z < -5000 m, stratification is thus considerable that turbulence values are found mainly below threshold value. For z < -6000 m, the stratification becomes weak and the buoyancy frequency N computed over 100 dbar pressure difference reaches a minimum of N = 1±0.6 cpd (short for cycles per day) = 2.5±1.5f, where f denotes the local inertial frequency. Within the error bounds, but computed over several thousands of meters vertically -10000 < z < -7000 m, the 2002-data yield N = 0.75±0.1 cpd and the 2016-data yield N = 1.0±0.1 cpd. Despite the different stratification, the vertical turbulence fluxes, which are proportional to the turbulence dissipation rates, give comparable results for the two CTD-casts (Fig. 6b). Averaged over $-10000 < z < -7000$ m, we find $\varepsilon_{2002} = 1.9 \pm 1 \times 10^{-10}$ m$^2$ s$^{-3}$ for the 2002-data and $\varepsilon_{2016} = 1.8 \pm 1 \times 10^{-10}$ m$^2$ s$^{-3}$ for the 2016-data. Over this vertical range, these values are not significantly different from each other and from the values calculated from uncorrected 2016-data that, however, gave negligible values for $z < -10400$. Conductivity correction of the 2016-data yielded the near-bottom range [10200 10800] m average of $\varepsilon_{2016} = 5 \pm 3 \times 10^{-11}$ m$^2$ s$^{-3}$.

The equivalent turbulent flux values between the profiles and the different stratification result in approximately one-and-a-half-times larger turbulent diffusivity values for the 2002-data compared with the 2016-data (Fig. 6c). Like for the turbulence dissipation rate (~ turbulent flux), largest values are found in the range $-9000 < z < -7000$ m, but non-negligible values are found close to the trench-floor. Mean values for the $10000 < z < -7000$ m range of 2016-data are $K_z = 6 \pm 4 \times 10^{-3}$ m$^2$ s$^{-1}$ when computed using the Osborn (1980) parameterization. Mean values are higher using the parametrization suggested for estuaries under 'high' buoyancy Reynolds number values (Holleman et al. 2016), and tenfold less $K_z = 6 \pm 4 \times 10^{-4}$ m$^2$ s$^{-1}$ using the buoyancy Reynolds number $Re_b$ parametrization suggested for lakes with $Re_b < 400$ by Bouffard and Boegman (2013). In the present data $10^4 < Re_b = \varepsilon/\nu N^2 < 10^5$, which is generally considered high.

## 4. Discussion and conclusions

The present parameterization on mixing efficiency and turbulent diffusivity values leads to considerable ambiguity as demonstrated from our data above. As noted previously (van Haren et al. 2017), one cannot establish from CTD-profiles the particular type of turbulence and hence the mixing efficiency without knowledge on, e.g., the extent of the inertial subrange of shear-induced turbulence. One needs more extensive data sets, for example from moored instrumentation, to establish such knowledge. However, the present data can be used to establish some knowledge on the vertical variation of turbulent fluxes via turbulence dissipation rate calculations. The results between the two differently obtained data sets 14 years apart are consistent and show a weak turbulent vertical fluxing throughout the trench depth range down to the trench-floor. Values are comparable with or just below open ocean values (e.g., Gregg 1989). These turbulence values are about one order of magnitude larger than observed in the deep North-Pacific where microstructure profiler data gave $[\varepsilon] \sim 10^{-11}$ m$^2$ s$^{-3}$ around z = -5000 m (Yasuda et al. submitted). This is consistent with the below-threshold values found here between -6000 < z < -5000 m. It remains to be established whether the observed larger turbulence activity in deep trench waters is associated with enhanced carbon fluxes reported by Glud et al. (2013).

The water sample based polynomial correction to the conductivity data provides a consistent weakly stable deep trench stratification. This improves Mantyla and Reid's (1978) presentation of a nearly 4000 m tall unstable water column and which was attributable to their zero change in salinity values between their 7000 and 11000 dbar sampling levels. While temperature and salinity both positively contribute to vertical density variations, the contributions of salinity dominate in the present data. For z < -7000 m a tight temperature – density relationship is found, with <10% difference between coefficients for the 2002- and 2016-data. This results in <15% difference in mean turbulence dissipation rate values, and <7% in eddy diffusivity, which is well within error bounds.

Our proposed correction for artificial pressure effects on conductivity-data by applying a mathematical polynomial fit across the entire water column has no physical foundation. It is obvious that a linear trend is not observed, due to temperature contributions to conductivity. We hypothesize that the ceramics of the TC-duct have a small nonlinear response in addition to the linear pressure functionality. With the successful application of the additional conductivity correction to the 2016-Sonne data we speculate that the (A1) polynomial may work for other hadal depth SBE911 data sets, also from other trenches.

The corrected salinity data still weakly increase with depth, together with a decrease in temperature, which may point at an influx of dense modified Antarctic bottom water. For improved establishing such water mass transformations we require future extensive deep CTD-sampling.

**Acknowledgements**

We thank the masters and crews of the R/V Hakuho Maru and R/V Sonne for the pleasant cooperation during the operations at sea. We acknowledge Dr. K. Taira who planned the CTD-observations in the Mariana Trench during the Hakuho Maru Cruise in 2002.

**Appendix Correction coefficient**

Additional pressure correction to Sea-Bird SBE-911 Conductivity data at hadal depths.

$$Cc(p) = Co(p) - a_0 - a_1 p^1 - a_2 p^2 - a_3 p^3 - a_4 p^4 - a_5 p^5 - a_6 p^6 - a_7 p^7 - a_8 p^8 - a_9 p^9 \ [S\ m^{-1}], \qquad (A1)$$

with best-fit coefficients,

$a_0 = -1.0940527 \times 10^{-03}$

$a_1 = +2.0797759 \times 10^{-06}$

$a_2 = -1.9887985 \times 10^{-09}$

$a_3 = +1.0474409 \times 10^{-12}$

$a_4 = -3.2735743 \times 10^{-16}$

$a_5 = +6.3408610 \times 10^{-20}$

$a_6 = -7.7038164 \times 10^{-24}$

$a_7 = +5.7145710 \times 10^{-28}$

$a_8 = -2.3640827 \times 10^{-32}$

$a_9 = +4.1779238 \times 10^{-37}$.

Here, Cc is the corrected conductivity and Co the original data from (2). Formula (A1) is applicable for all pressures, but yields noticeable corrections for great depths only.

**Fig. 1**. Sites of the 2002-Hakuho Maru (white) and 2016-Sonne (black) CTD stations in the Challenger Deep, Mariana Trench, North-Pacific. One minute grid version of the ocean topography database presented by Smith and Sandwell (1997). Local calibrated multibeam echosounder depth data are given in [m], which may be compared with the database values of -10945 and -10377 m for the 2016- and 2002-sites, respectively.

**Fig. 2**. Lower 5500 m of the 2002-Hakuho Maru and 2016-Sonne CTD profiles from the Challenger Deep, Mariana Trench. Data without corrections except for ship-motions including 0.05 cps, cycle per second, low-pass filtering, see text. (a) Conservative Temperature. (b) Absolute Salinity. (c) Potential density anomaly, referenced to 11000 dbar.

**Fig. 3**. Magnification of 2002-data in Fig. 2 demonstrating a characteristic deep trench instability around z = -7650 m. The black bars indicate the approximate error bars for the 0.05 cps low-pass filtered data.

**Fig. 4**. As Fig. 2b, but for corrections to 2002-CTD data using water sample and laboratory information. Uncorrected data in green, corrected data in black, water sample laboratory data in purple (x). The spike of about 0.01 g kg$^{-1}$ at 6800 m is an apparently extreme noise contamination in the water sampling.

**Fig. 5**. Comparison between original and corrected data and the polynomial fit to the corrections. (a) Entire profile of conductivity difference between original and water-sample-corrected data (black) with a 9$^{th}$ order polynomial fit (purple). (b) As a., but for the lower 5500 m of Absolute Salinity difference between water-sample-corrected data and data computed from the polynomial fit-profile of a. (black). In purple the difference between water-sample and electronic CTD-data during water sample taking. In blue the difference between original (0.05 cps low-pass filtered) data and polynomial fit-data from the 2016-Sonne cruise. (c) Lower 5500 m profile of Absolute Salinity of: 2002-data corrected using water samples (black), 2002-data corrected using $9^{th}$ order polynomial on conductivity difference in a. (red), 2016-data using $9^{th}$ order polynomial of 2002-Hakuho-Maru -conductivity correction (blue).

**Fig. 6**. Lower 5500 m of turbulence characteristics computed from $9^{th}$ order polynomial 2002-Hakuho Maru-conductivity corrected downcast 2002- and 2016-data applying a threshold of $7\times10^{-5}$ kg m$^{-3}$. (a) Density anomaly referenced to 11000 dbar. (b) Logarithm of dissipation rate computed from the profiles in a., averaged over 200 m vertical intervals. Values are zero when threshold is not passed. (c) As b., but for eddy diffusivity. The dashed profiles indicate values using the parameterization proposed for lake data by Bouffard and Boegman (2013).

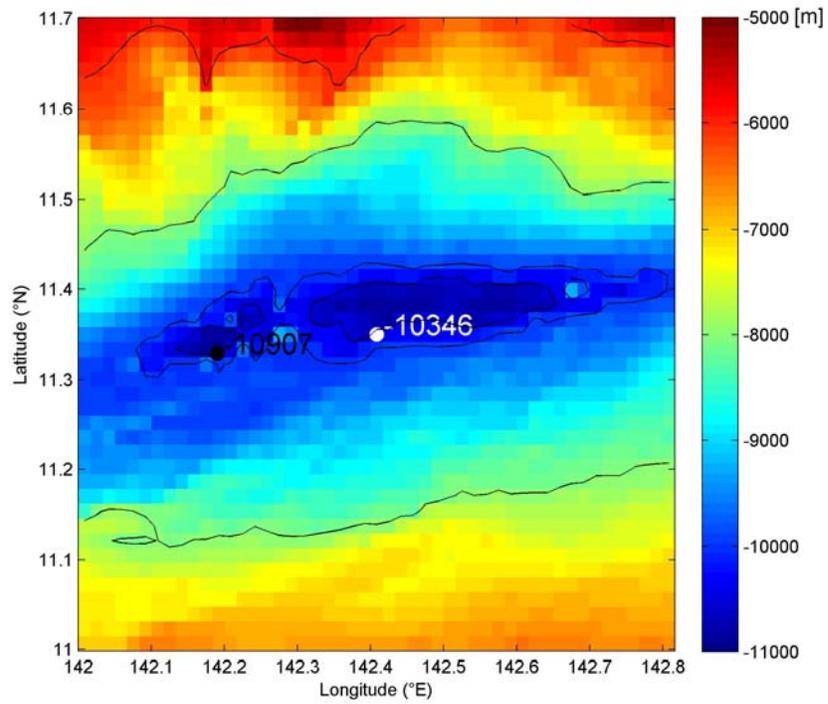

**Fig. 1**. Sites of the 2002-Hakuho Maru (white) and 2016-Sonne (black) CTD stations in the Challenger Deep, Mariana Trench, North-Pacific. One minute grid version of the ocean topography database presented by Smith and Sandwell (1997). Local calibrated multibeam echosounder depth data are given in [m], which may be compared with the database values of -10945 and -10377 m for the 2016- and 2002-sites, respectively.

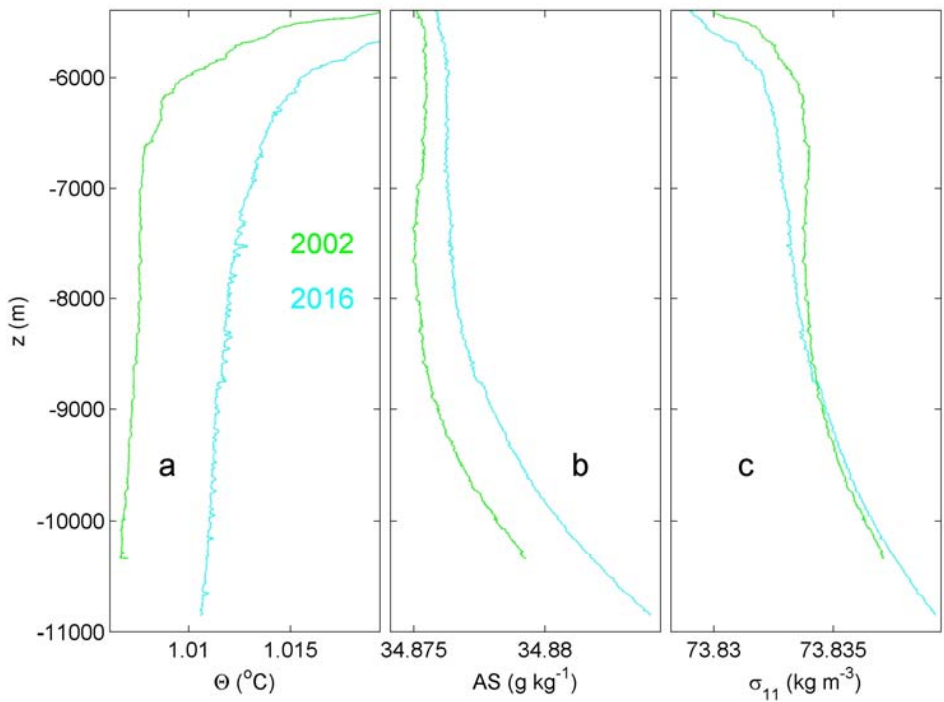

**Fig. 2**. Lower 5500 m of the 2002-Hakuho Maru and 2016-Sonne CTD profiles from the Challenger Deep, Mariana Trench. Data without corrections except for ship-motions including 0.05 cps, cycle per second, low-pass filtering, see text. (a) Conservative Temperature. (b) Absolute Salinity. (c) Potential density anomaly, referenced to 11000 dbar.

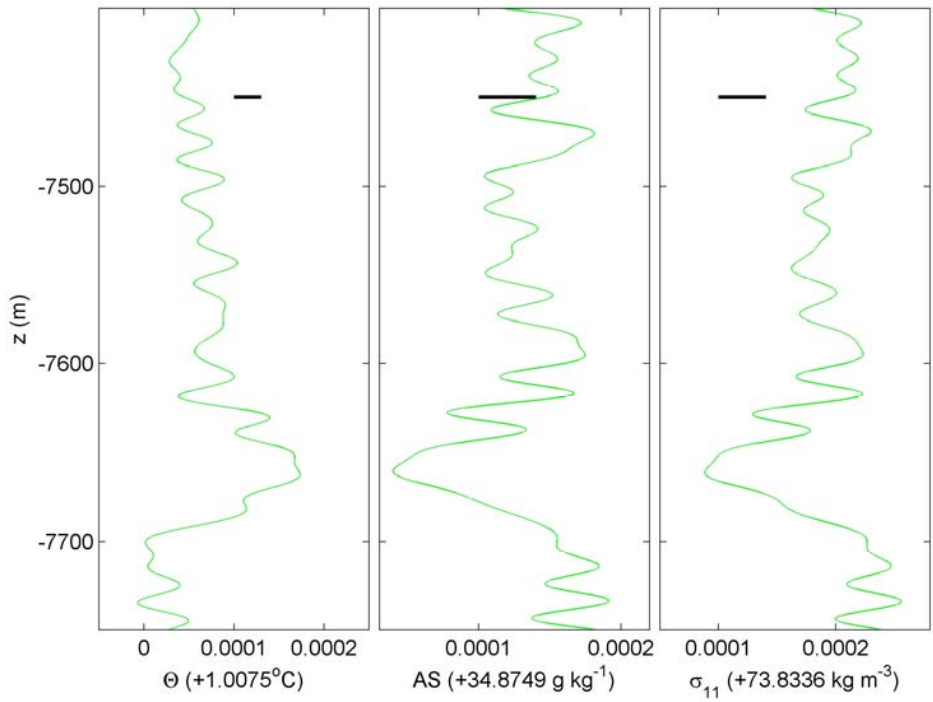

**Fig. 3**. Magnification of 2002-data in Fig. 2 demonstrating a characteristic deep trench instability around z = -7650 m. The black bars indicate the approximate error bars for the 0.05 cps low-pass filtered data.

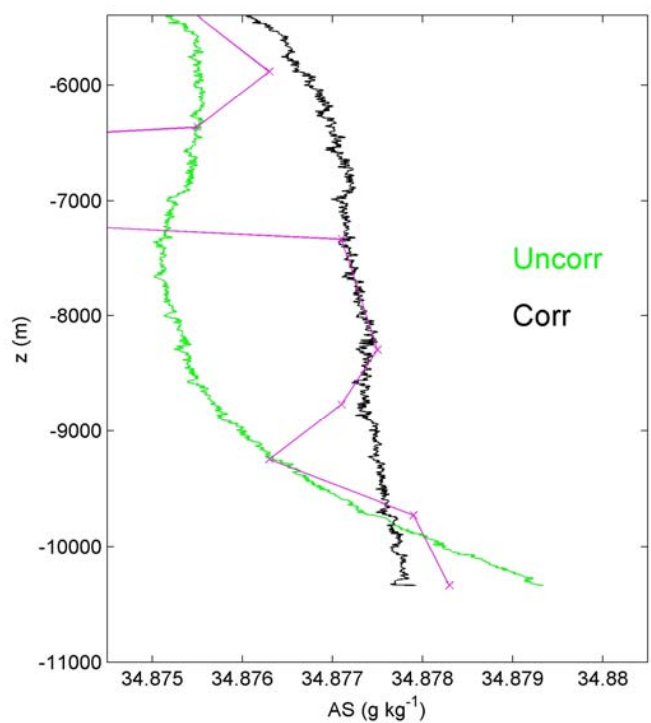

**Fig. 4**. As Fig. 2b, but for corrections to 2002-CTD data using water sample and laboratory information. Uncorrected data in green, corrected data in black, water sample laboratory data in purple (x). The spike of about 0.01 g kg$^{-1}$ at 6800 m is an apparently extreme noise contamination in the water sampling.

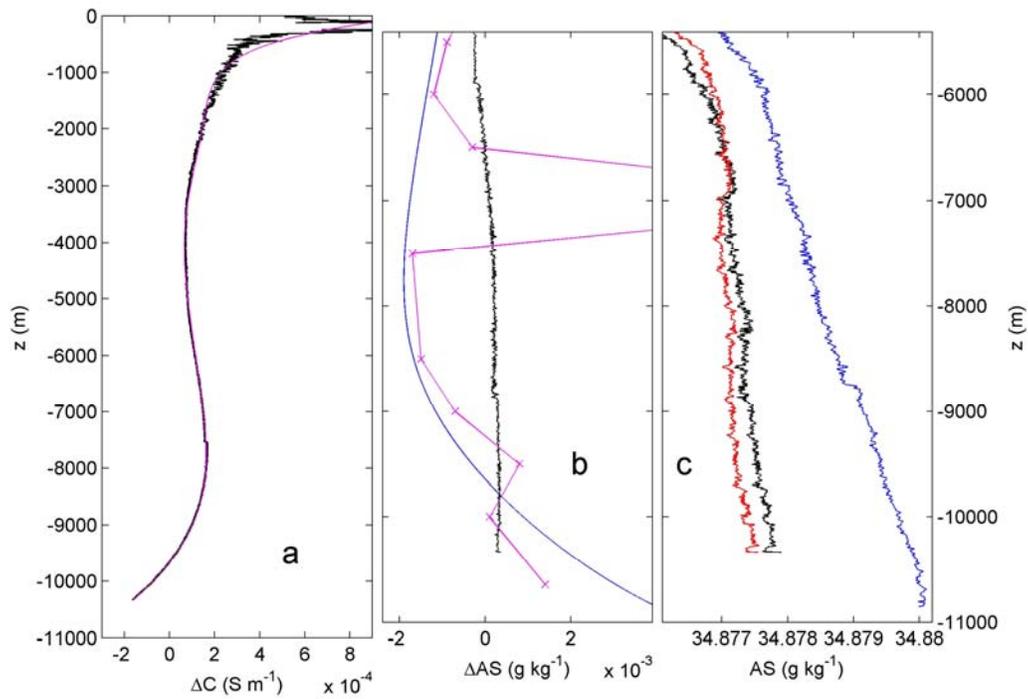

**Fig. 5**. Comparison between original and corrected data and the polynomial fit to the corrections. (a) Entire profile of conductivity difference between original and water-sample-corrected data (black) with a 9$^{th}$ order polynomial fit (purple). (b) As a., but for the lower 5500 m of Absolute Salinity difference between water-sample-corrected data and data computed from the polynomial fit-profile of a. (black). In purple the difference between water-sample and electronic CTD-data during water sample taking. In blue the difference between original (0.05 cps low-pass filtered) data and polynomial fit-data from the 2016-Sonne cruise. (c) Lower 5500 m profile of Absolute Salinity of: 2002-data corrected using water samples (black), 2002-data corrected using 9$^{th}$ order polynomial on conductivity difference in a. (red), 2016-data using 9$^{th}$ order polynomial of 2002-Hakuho-Maru -conductivity correction (blue).

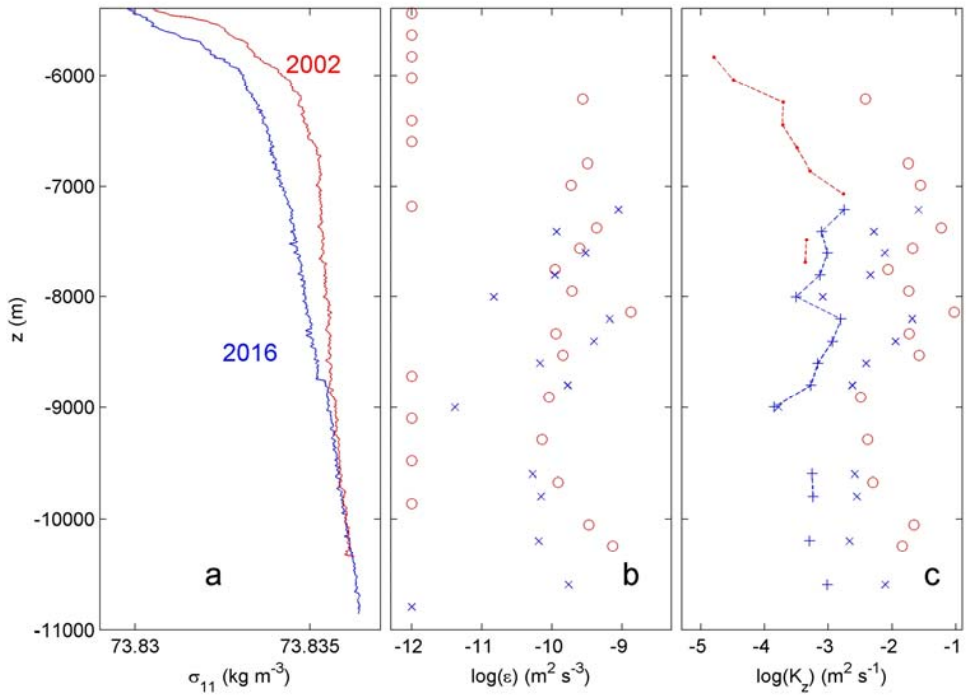

**Fig. 6**. Lower 5500 m of turbulence characteristics computed from 9$^{th}$ order polynomial 2002-Hakuho Maru-conductivity corrected downcast 2002- and 2016-data applying a threshold of $7 \times 10^{-5}$ kg m$^{-3}$. (a) Density anomaly referenced to 11000 dbar. (b) Logarithm of dissipation rate computed from the profiles in a., averaged over 200 m vertical intervals. Values are zero when threshold is not passed. (c) As b., but for eddy diffusivity. The dashed profiles indicate values using the parameterization proposed for lake data by Bouffard and Boegman (2013).